----------
X-Sun-Data-Type: default
X-Sun-Data-Name: certo.tex
X-Sun-Charset: iso-8859-1
X-Sun-Content-Lines: 241

\documentstyle[12pt]{article}

\centerline{\bf  {\Large Anti-symmetric Tensor Matter Fields}}
\centerline {\bf{\Large and Non-Linear $\sigma$-Model}}
\vspace{0.7cm}
\centerline{M.G.Negr\~ao$^{\dag}$\footnote{negrao@cat.cbpf.br},A.Penna-Firme$^{(**)\dag}$\footnote{andrepf@cat.cbpf.br} and}
\centerline{J.A.Helay\"el-Neto$^{(*)\dag}$\footnote{helayel@cat.cbpf.br}}
\vspace{.5cm}
\centerline{$^{\dag}$Centro Brasileiro de Pesquisas F\'{\i} sicas- CBPF/CNPq }
\centerline {$^{(*)}$Universidade Cat\'olica de Petr\'opolis}
\centerline{$^{(**)}$Faculdade de Educa\c{c}\~ao da Universidade Federal do Rio de Janeiro\\}

\begin{document}

\hspace\parindent

\begin{abstract}
\vspace{0.3cm}

{\footnotesize{\bf The equivalence between rank-$2$ anti-symmetric
tensor fields,  considered as  gauge potentials, and torsionless non-linear
$\sigma$-
models suggests us to study the possibility of coupling tensorial
matter with Yang-Mills fields, through the gauging of the isometries
of the target space.
 We show that this coupling is actually possible; however,
the matter appears no longer as an
elementary field, but rather as a composite one, expressed in
terms of the bosonic degrees of freedom
of the $\sigma$-model.
A possible phenomenological application is presented that describes
the interactions among vector mesons
in terms of the geometrical properties of the target manifold.
Also, spin-2 meson resonances may naturally be
accommodated whenever the $\sigma$-model's target manifold is
non-symmetric.}}
\end{abstract}

\newpage

\section{Introduction}

It is well-known that anti-symmetric tensor fields, $B_{\mu \nu}$,
subject to the gauge transformations
$\delta  B_{\mu \nu} = \partial_{\mu} \zeta_{\nu} - \partial_{\nu} \zeta_\mu$,
are equivalent to massless scalar particles \cite{{1},{1.2},{1.3},{1.4}}. Starting from the
Abelian theory, we can iterate consistency conditions to build up the non-Abelian version where self-interactions terms will be present.

 In a first-order formulation of this tensor gauge theory, it is easy to show that there is an equivalence between these non-Abelian tensor fields and a non-linear $\sigma$-model without torsion\cite{town}. Treated as gauge fields, it is already known\cite{h} that a no-go theorem prevents the construction of a gauge theory whereby usual spin-1 gauge bosons interact with anti-symmetric tensor potential according to the Yang-Mills prescrition.

In this work, it is our aim to consider anti-symmetric tensor fields as matter, rather than as gauge fields. With this purpose in mind, it is possible to write down the action for a non-linear $\sigma$-model with gauged isometries. We find thereby an expression for the coupling of tensorial matter (equivalent to the scalar fields of the $\sigma$-model) with Yang-Mills gauge fields defined over a coset-like manifold. We get that this coupling implies a non-elementary character for the tensorial field: the tensorial matter is, then, written in terms of the degrees of freedom of the $\sigma$-model.

Following this procedure, we verify that these matter fields might be introduced as sources of torsion on target manifold. Depending on the isometries of this space, we are able to construct a phenomenological approach in which the interactions between multiplets of vector mesons are understood in terms of the geometrical properties of the torsion generated by the tensorial fields of matter.
Vector and spin-2 resonances are associated to composite objects
written down in terms of the connection, torsion and curvature.

\section{The Geometry of the $\sigma$-Model}

The action of the non-linear $\sigma$-model is given by\cite{2}
\begin{eqnarray}
S=\int d^4x \frac{1}{2} \partial_{\mu} {\theta}^m \partial^{\mu} {\theta}^m
\end{eqnarray}

The set $({\theta}^1,{\theta}^2,...,{\theta}^N)$ defines the coordinates of the $\sigma$-model with a metric $g_{mn}(\theta)$. These fields fulfill, for example,  a non-linear constraint:
\begin{eqnarray}
{\theta}_1^2+{\theta}_2^2+...+{\theta}_N^2=1.
\end{eqnarray}
This constraint induces the mapping ${\cal M}^{1,3} \to S^{N-1}$. The global symmetry is $SO(N)$, and it is straightfoward to obtain the metric of this manifold:
\begin{eqnarray}
g_{mn}(\theta)=\delta_{mn}-\frac{\theta_m\theta_n}{1-\theta_m\theta_m}.
\end{eqnarray}

Consequently, we might rewrite our original action:
\begin{eqnarray}
S=\int d^4x \frac{1}{2}g_{mn}(\theta)\partial_\mu\theta^m\partial^\mu\theta^n.
\end{eqnarray}

The coset space in this case is $S^{N-1}=\frac {SO(N)}{SO(N-1)}$, and the Maurer-Cartan equations define quantities like curvature and torsion on this manifold:
\begin{eqnarray}
&de+e\wedge e=0,\\
&de+w \wedge e =T,\\
&dw+w \wedge w =R.
\end{eqnarray}a

The connection is given by:
\begin{eqnarray}
\omega_m^{ab}=\frac{1}{2}{\Omega_m^{ab}}+\frac{1}{2}e_m^c{K_c^{ab}},
\end{eqnarray}
where $\Omega$ and $K$ are, respectively, the non-holonomicity
coefficients and the contortion.

We may obtain the equivalence with the $\sigma$-model by identifying the pull-back of the connection with the gauge fields:
\begin{eqnarray}
V_\mu^{ab}=\omega_m^{ab}\partial_\mu \theta^m \rightarrow F_{\mu\nu}^{ab}=R_{mn}^{ab}\,\,\,\partial_\mu \theta^m \partial_\nu\theta^n,
\end{eqnarray}
where $R_{mn}^{ab}(\theta)$ is the Riemann curvature tensor of the manifold of $\sigma$-model, and may be found by the second Maurer-Cartan equation:
\begin{eqnarray}
R_{mn}^{ab}=\partial_m\omega_n^{ab}-\partial_n\omega^{ab}_m+\omega_m^{ac}\omega_{nc}^b-\omega_n^{ac}\omega_{mc}^b.
\end{eqnarray}

The 2-form torsion of the manifold is given by the first Maurer-Cartan equation:
\begin{eqnarray}
T_{mn}^a=\frac{1}{2}\biggl (\partial_me_n^a-\partial_ne_m^a+\omega_{mb}^ae^b_n-\omega_{nb}^ae_m^b \biggr ).
\end{eqnarray}
In order to introduce the 2-forms fields as matter degree of freedom, we need torsion, mainly because it is a tensor and it is pull-back defines a rank-2 field over space-time:
\begin{eqnarray}
W_{\mu\nu}^a=T_{mn}^a(\theta)\partial_\mu\theta^m\partial_\nu\theta^n.
\end{eqnarray}

As we see, $T_{mn}^a$ is a real tensor, so,  $W_{\mu\nu}^a$ is an anti-symmetric tensor matter field, which in principle may describe massless spin-0 or massive spin-1 particles.

\section{Connection with Non-Linear $\sigma$-Model}

In our approach, we propose the following identification:
\begin{eqnarray}
W_{\mu\nu}^a &=& T_{mn}^a(\theta)\partial_\mu\theta^m\partial_\nu\theta^n, \\
F_{\mu\nu}^a(V) &=& R_{mn}^{ab}\partial_\mu\theta^m\partial_\nu\theta^n, \\
V_\mu^{ab} &=& \omega_m^{ab} \partial_\mu \theta^m, \\
\end{eqnarray}
where $V_{\mu}^{ab}$ acts an Yang-Mills connection and $W_{\mu\nu}^a$ transforms as spin-0 or spin-1 matter fields.

As it can be seen, these matter fields are no longer fundamental ones, but composite particles, written in terms of the coordinates of the $\sigma$-model, the curvature and the torsion. This fact is in agreement with the no-go theorem stated in the work of reference\cite{h}. In this sense, this model could serve only as an effective one describing some low-energy physics. A good aplication of this effective model might be the interactions among mesons: not only scalar mesons, as in the standard non-linear $\sigma$-model approach, but also vector particles. In this scheme, the gauge field could be the quantum of this effective interaction, which is also a composite field.

\section{Manifolds G/H with Torsion}

To present a concrete application, it is crucial to find manifolds of the form {G/H}, such that the embedding of H into G allows for non-trivial torsion.

The splitting of the generators, which specify the embendding of H
into G, is given as follows\cite{3}
\begin{eqnarray}
\begin{array}{c}
[Q_i,Q_j]=f_{ijk}Q_k, \\

[Q_i,Q_a]=f_{ia}^b Q_b, \\

[Q_a,Q_b]=f^i_{ab}Q_i+{f^c_{ab}Q_c},
\end{array}
\end{eqnarray}
where
\begin{eqnarray}
\begin{array}{c}
T^a=kf^a_{bc}e^b \wedge e^c, \\
\omega^a_b=-f^a_{bi}e^i-(1+k)f^a_{bc}e^c.
\end{array}
\end{eqnarray}
The generators $Q_{i}$ lie in the subalgebra $H$ of $G$, whereas
the $Q_{a}$'s stand for the generators that are in $G$ but not in
$H$.

\section{Contact With Phenomenology}

In order to accommodate multiplets of vector mesons, we needed to search for coset spaces with a non-symetric embedding. All the components of a given  meson multiplet must have approximately the same mass:
\begin{enumerate}
\item $\frac{G_2}{SU(3)}$:

$6$-dimensional manifolds $\longrightarrow$ $6$-plet: $a_1(1270 GeV)$ and $b_1(1285 GeV)$

\item $\frac{SO(4)}{SU(2)}$:

$3$-dimensional manifolds $\longrightarrow$ $3$-plet: $\rho (770 GeV)$

\item $\frac{SU(3)}{U(1)}$ and $\frac{Sp(4)}{SU(2)}$:

$7$-dimensional manifolds$\longrightarrow$ $7$-plet: $a_1(1270GeV), b_1(1285GeV),\\
f_1(1285GeV)$.
\end{enumerate}
{\bf All these coset manifolds (G/H) have a torsion induced by the non-symmetric embedding of H into G}. As a massive fields, {\it skew-symmetric} the rank-2 tensor built up from torsion describes spin-1 mesons. The almost-degenerate resonances appear in multiplets whose degeneracy is given by the discusion of the coset spaces above.

\section{On Spin-2 Resonances}

With the geometrical objects we have at our disposal, we wish now to propose a possible description for the almost-degenerate spin-2 resonances of the meson spectrum. Symmetric rank-2 tensor fields associated to massive spin-2 particles can be identified by the following composite fields:
\begin{eqnarray}
S_{\mu\nu}={T_{i}}_{k}^{a}
{T_j}^{k}_{a}\partial_{\mu}\theta^i\partial_{\nu}\theta^j,
\end{eqnarray}
to the singlet case, and
\begin{eqnarray}
S_{\mu\nu}^a=R_{(i}^{kab}T_{j)}^{kb}\partial_{\mu}\theta^{i}\partial_{\nu}\theta^{j}=S_{\nu\mu}
\end{eqnarray}
for non-trivial multiplets. However, a close inspection of the
meson summary table indicates the appearance of singlets, doublets, 3-plets and 6-plets of spin-2 resonances. So, the
non-symmetric spaces $G_{2}/SU(3)$ and $SO(4)/SU(2)$ may well set
the geometry for the 3-plet and 6-plet of spin-2 mesons. Again,
torsion seens to be the key element for the construction of
composite fields describing resonances of non-trivial spin.

\section{Conclusion}

This work shows that  non-linear $\sigma$-models with torsion defined on a coset space, $G/H$, with a non-symmetric embedding allow for the introduction of non-Abelian tensor fields that behave like massive spin-1 matter fields. These anti-symmetric fields are introduced as sources for torsion in the manifold of the $\sigma$-model. As a consequence, it is shown that multiplets of vector and spin-2 mesons can be introduced in such a way that they interact with scalar mesons in the framework of an effective (non-renormalisable) theory described by the $\sigma$-model.

In this context, we claim that not only scalar mesons can be effectively described by non-linear $\sigma$-models, but also the vector ones. For this we must introduce torsion in the manifold.

\end{document}